\documentclass[aps,graphicx,10pt,twocolumn]{revtex4}
\usepackage{amssymb}
\usepackage{amsmath}
\usepackage{amscd}
\usepackage{subfigure}
\usepackage{graphicx}

\begin{document}

\title[Short Title]{Using shortcut to adiabatic passage for the ultrafast quantum state transfer in cavity QED system}

\author{Mei Lu$^{1}$}
\author{Li-Tuo Shen$^{1}$}
\author{Yan Xia$^{1,}$\footnote{E-mail: xia-208@163.com}}
\author{Jie Song$^{2}$}

\affiliation{$^{1}$Lab of Quantum Optics, Department of Physics,
Fuzhou University, Fuzhou 50002, China\\$^{2}$Department of Physics,
Harbin Institute of Technology, Harbin 150001, China}
\begin{abstract}
We propose an alternative scheme to implement the quantum state
transfer between two three-level atoms based on the invariant-based
inverse engineering in cavity quantum electronic dynamics (QED)
system. The quantum information can be ultrafast transferred between
the atoms by taking advantage of the cavity field as a medium for
exchanging quantum information speedily. Through designing the
time-dependent laser pulse and atom-cavity coupling, we send the
atoms through the cavity with a short time interval experiencing the
two processes of the invariant dynamics between each atom and the
cavity field simultaneously. Numerical simulation shows that the
target state can be ultrafast populated with a high fidelity even
when considering the atomic spontaneous emission and the photon
leakage out of the cavity field. We also redesign a reasonable
Gaussian-type wave form in the atom-cavity coupling for the
realistic experiment operation.
\end{abstract}

\maketitle

\section{INTRODUCTION}

For various applications ranging from quantum storage and quantum
communication \cite{PRL-96-080501-2006,Nature-488-185-2012},
reliable quantum state transfer (QST) between two qubits has become
an essential ingredient in the quantum information processing
\cite{Science-261-5128-1993}. Therefore, there is much interest in
the QST for recent years
\cite{PRA-71-023805-2005,PRA-72-012339-2005,PRA-76-062304-2007,
PRA-82-054303-2010,PRL-106-040505-2011,PRL-108-153603-2012}.
Experimentally, QST has been demonstrated with superconducting phase
qubits and transmon qubits in the cavity quantum electronic dynamics
(QED) system \cite{Nature-449-438-2007,Nature-449-443-2007}. Several
theoretical proposals based on the technique of adiabatic passage
have been applied in the implementation of the QST, and these
adiabatic passage techniques include the stimulated Raman adiabatic
passage (STIRAP)
\cite{PRA-40-6741-1989,RMP-70-1003-1998,RMP-79-53-2007} and the
fractional stimulated Raman adiabatic passage (f-STIRAP)
\cite{JPB-32-4535-1999,PRA-71-023805-2005}. For example,
Amniat-Talab \emph{et al.} \cite{PRA-40-6741-1989} used the
techniques of STIRAP and f-STIRAP to successfully transfer the
quantum state between the $\Lambda$-type atoms and the photons.
Although the adiabatic passage techniques are robust against the
fluctuations of experimental parameters, the operation time needed
to complete the QST is rather long in most cavity QED systems.
However, the direct atom-photon interactions typically decays with
the evolution time leading to the limitation for the perfect QST by
adopting the adiabatic passage techniques.

Recently, Chen and Muga \cite{PRA-86-033405-2012} achieved the fast
population transfer within two internal states of a single
$\Lambda$-type atom by the invariant-based inverse engineering, and
only two resonant laser pulses were used. The invariant-based
inverse engineering combines the advantages of resonant pulses whose
operation time is short and adiabatic technique which is robust
against the variation of parameters, which has been used for
different systems
\cite{PRL-104-063002-2010,PRA-83-062116-2011,EPL-93-23001-2011,NJP-13-113017-2011,
PRA-83-013415-2011,PRA-84-043415-2011,OL-37-5118-2012,JPSJ-81-024007-2012,
NJP-14-013031-2012,PRA-83-043804-2011,PRA-84-031606R-2011,PRA-84-051601R-2011},
except for the cavity QED system. The ultrafast population transfer
between two or more atoms is a fundamental operation for scalable
quantum information processors. However, previous studies based on
the invariant-based inverse engineering focus on the ultrafast
population transfer within two internal states of a single atom, and
has not been devoted to the ultrafast population transfer between
two or more atoms. Here, the major obstacle for the ultrafast
population transfer between two or more atoms through the
invariant-based inverse engineering exists in finding an appropriate
medium that can be used to exchange energy or information with the
atoms speedily.

To improve the efficiency of QST based on the traditional adiabatic
passage in cavity QED system, we propose an alternative scheme based
on the invariant-based inverse engineering to implement the
ultrafast quantum state transfer between two $\Lambda$-type atoms in
this paper. We take advantage of the cavity field as a medium for
exchanging information between the atoms speedily, which is very
different from that in Ref. \cite{PRA-86-033405-2012} where the
population transfer is confined in two internal states of a single
atom. Consider the photon leakage out of the cavity, the atoms are
sent through the cavity with a short time interval, which suffer the
oppositive variation tendency in the time-dependent laser pulse and
atom-cavity coupling, i.e., two processes of the invariant-based
inverse engineering between each single $\Lambda$-type atom and the
cavity field happen simultaneously. We find that the operation time
needed to complete the QST is short enough before a photon leaks out
of the cavity, and this is an obvious improvement compared with the
QST based on the traditional adiabatic passage, resulting in the
target state with a high fidelity even when taking the system's
decoherence into consideration, including the atomic spontaneous
emission and the photon leakage out of the cavity. We also redesign
the shape from a sinusoidal-wave form to a Gaussian-wave form in the
time-dependent atom-cavity coupling strength for the realistic
experiment. Compared with the traditional QST proposals, our
improvement in the QST based on the invariant-based inverse
engineering is very feasible with the current cavity QED technology
\cite{PRL-87-037902-2001}, and the present idea can be generalized
to the QST models among three and more atoms.

\section{Invariant-based inverse engineering in the cavity QED system}

\subsection{Invariant dynamics between a single $\Lambda$-type atom and the cavity mode}
\begin{figure}
\centering
\includegraphics[width=0.4\columnwidth]{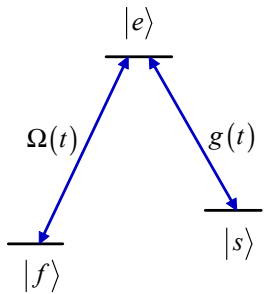}
\caption{ The level configuration for our setup. The transition
$|e\rangle\leftrightarrow|s\rangle$ couples resonantly to the cavity
mode with the time-dependent coupling coefficient $g(t)$ and a
classical laser drives the transition
$|e\rangle\leftrightarrow|f\rangle$ with the time-dependent coupling
coefficient $\Omega(t)$.} \label {Fig.1}
\end{figure}
We generalize the dynamics of invariant-based engineering to the
cavity QED system containing a $\Lambda$-type atom and the cavity
mode. In the atom-cavity system, as shown in Fig. 1, the single
$\Lambda$-type atom has an excited state $|e\rangle$ and two ground
states $|f\rangle$ and $|s\rangle$. The transition
$|e\rangle\leftrightarrow|f\rangle$ is resonantly driven by the
laser pulse with the time-dependent Rabi frequency $\Omega(t)$ and
the transition $|e\rangle\leftrightarrow|s\rangle$ resonantly
couples to the cavity mode $a$ with the time-dependent coupling
coefficient $g(t)$. $\Omega(t)$ and $g(t)$ are assumed to be real in
the following for simplicity. The state vector $|M,N\rangle$ denotes
that the atom is in the state $|M\rangle\ (M=f,s,e)$ and there are
$N\ (N=0,1,2,...)$ photons in the cavity field. Note that the total
excitation of the atom-cavity system is conserved during the state
evolution. In the interaction picture, the time-dependent
Hamiltonian $H(t)$ under the rotating-wave approximation is block
diagonal in the $(N+1)$-excitation subspace
$\{|f,N\rangle,|e,N\rangle,|s,N+1\rangle\}$. The vector
$|s,0\rangle$ here is not coupled to any other ones, therefore, one
can thus restrict the problem to the projection of the Hamiltonian
in the single-excitation subspace
$\{|f,0\rangle,|e,0\rangle,|s,1\rangle\}$ as following $(\hbar=1)$:
\begin{eqnarray}\label{e1}
H(t)=\left({\begin{array}{*{20}{c}}
{0}&{\Omega(t)}&{0}\\
{\Omega(t)}&{0}&{g(t)}\\
{0}&{g(t)}&{0}
\end{array}}\right).
\end{eqnarray}
The instantaneous eigenstates of $H(t)$ are
$|n_0(t)\rangle=[\cos\theta,\ 0,\ -\sin\theta]^T$ and
$|n_\pm(t)\rangle=\frac{1}{\sqrt{2}}[\sin \theta,\ \pm1,\
\cos\theta]^T$, with the corresponding eigenvalves $E_0=0$ and
$E_{\pm}$ $=\pm\omega$, where $\theta=\arctan[\Omega(t)/g(t)]$ and
$\omega=\sqrt{\Omega^2(t)+g^2(t)}$. When the adiabatic condition
$|\dot{\theta}|\ll|\omega|$ is satisfied and $\Omega(t)$ and $g(t)$
are applied through a counter-intuition passage as that in the
STIRAP, we can adiabatically transfer the population from the
initial state $|f,0\rangle$ to the final state $|s,1\rangle$ along
the dark state $|n_0(t)\rangle$. The key point here is to  speed up
the population transfer $|f,0\rangle$ $\rightarrow$ $|s,1\rangle$ in
the atom-cavity system by using the dynamics of invariant-based
inverse engineering. For the Hamiltonian in Eq. (\ref{e1}) that
possesses the SU(2) dynamical symmetry, an invariant Hermitian
operator can be employed to construct the invariant dynamics for our
atom-cavity system with a time-independent expectation value
$\langle I(t)\rangle$, which satisfies
\cite{PRA-86-033405-2012,PRA-53-3691-1996}:
\begin{eqnarray}\label{e2}
i\frac{\partial}{\partial t}I(t)-[H(t),I(t)]=0,
\end{eqnarray}
where the complex conjugate $I^{\dagger}(t)=I(t)$. The invariant
$I(t)$ is given by \cite{PRA-86-033405-2012}:
\begin{eqnarray}\label{e3}
I(t)=\mu\left({\begin{array}{*{20}{c}}
{0}&{\cos\gamma\sin\beta}&{-i\sin\gamma}\\
{\cos\gamma\sin\beta}&{0}&{\cos\gamma\cos\beta}\\
{i\sin\gamma}&{\cos\gamma\cos\beta}&{0}
\end{array}}\right),
\end{eqnarray}
where $\mu$ is an arbitrary constant with units of frequency to keep
$I(t)$ involving the energy dimension. The eigenstates of $I(t)$
with the corresponding eigenvalues $\lambda_0=0$ and
$\lambda_{\pm}=\pm1$ are respectively:
\begin{eqnarray}\label{e4}
|\Phi_0(t)\rangle=\left({\begin{array}{*{20}{c}}
{\cos\gamma\cos\beta}\\
{-i\sin\gamma}\\
{-\cos\gamma\sin\beta}
\end{array}}\right),
\end{eqnarray}
and
\begin{eqnarray}\label{e5}
|\Phi_{\pm}(t)\rangle=\frac{1}{\sqrt{2}}\left({\begin{array}{*{20}{c}}
{\sin\gamma\cos\beta\pm i\sin\beta}\\
{i\cos\gamma}\\
{-\sin\gamma\sin\beta\pm i\cos\beta}
\end{array}}\right).
\end{eqnarray}
According to Lewis Riesenfeld theory \cite{JMP-10-1458-1969}, the
general solution of the Schr\"{o}dinger equation with respect to the
instantaneous eigenstates of $I(t)$ can be written as:
\begin{eqnarray}\label{e6}
|\Psi(t)\rangle=\sum_{m=0,\pm}C_{m}e^{i\alpha_{m}}|\Phi_{m}(t)\rangle,
\end{eqnarray}
where $C_m$ is a time-independent amplitude and $\alpha_m$ is the
Lewis-Riesenfeld phase with the following form:
\begin{eqnarray}\label{e7}
\alpha_m(t^{'})=\int_0^{t^{'}}dt\langle\Phi_m(t)|\big[i\frac{\partial}{\partial
t}-H(t)\big]|\Phi_m(t)\rangle,
\end{eqnarray}
where $t^{'}$ is the total interaction time. And the time-dependent
parameters $\gamma(t)$ and $\beta(t)$ should satisfy the following
auxiliary equations \cite{PRA-86-033405-2012}:
\begin{eqnarray}\label{e8-e9}
\dot{\gamma}&=&\Omega\cos\beta-g\sin\beta,\\
\dot{\beta}&=&\tan\gamma(g\cos\beta+\Omega\sin\beta),
\end{eqnarray}
where the dot represents a time derivative.

\subsection{Fast population transfer for the atom-cavity system}

The explicit expressions of $\Omega(t)$ and $g(t)$ to be designed
can be inversely derived from Eqs. (8) and (9):
\begin{eqnarray}\label{e10-e11}
\Omega&=&\dot{\beta}\cot\gamma\cos\beta-\dot{\gamma}\sin\beta,\\
g&=&\dot{\beta}\cot\gamma\sin\beta+\dot{\gamma}\cos\beta.
\end{eqnarray}
For a single-mode driving, the atom-cavity system is assumed to be
prepared in one of the eigenstates ($|\Phi_m(t)\rangle$) of $I(t)$
initially, then the atom-cavity system is driven along this
instantaneous eigenstate $|\Phi_m(t)\rangle$ without worrying about
its transition to the other eigenstates, while the adiabatic
condition is unnecessary here. To achieve the fast population
transfer from the atom to the cavity field, the feasible parameters
$\gamma(t)$ and $\beta(t)$ can be chosen as
\cite{PRA-86-033405-2012}:
\begin{eqnarray}\label{e12-e13}
\gamma(t)&=&\epsilon,\\
 \beta(t)&=&\pi t/2t_f,
\end{eqnarray}
leading to:
\begin{eqnarray}\label{e14-e15}
\Omega(t)&=&(\pi/2t_f)\cot\epsilon\cos(\pi t/2t_f),\\
g(t)&=&(\pi/2t_f)\cot\epsilon\sin(\pi t/2t_f),
\end{eqnarray}
where $\epsilon$ is small value to be chosen later and
$\epsilon\neq0$.

However, when the initial state is $|f,0\rangle$, the atom-cavity
model is essentially a multi-mode driving rather than a single-mode
driving, meaning the time-dependent wave function
$|\Psi(t)\rangle=\chi_1(t)|f,0\rangle+\chi_2(t)|s,0\rangle
+\chi_3(t)|s,1\rangle$ involve contributions stemming from all the
eigenvectors of the invariant $I(t)$
\cite{PRA-86-033405-2012,OL-37-5118-2012}, where
$|\chi_1(t)|^2+|\chi_2(t)|^2+|\chi_3(t)|^2=1$,
$|\Psi(0)\rangle=|f,0\rangle$, and $|\Psi(t_f)\rangle=|s,1\rangle$.
The superposition of $|\Phi_0(0)\rangle$ and $|\Phi_{\pm}(0)\rangle$
corresponds to the initial bare state $|\Psi(0)\rangle$, while the
superposition of $|\Phi_0(t_f)\rangle$ and $|\Phi_{\pm}(t_f)\rangle$
corresponds to the final bare state $|\Psi(t_f)\rangle$. The present
setup based on the multi-mode driving provides the ultrafast
population transfer with less intensities of energy than that based
on the single-mode driving. The coefficients $\Omega(t)$ and $g(t)$
are chosen the same as those in the situation of the single-mode
driving, corresponding to Eqs. (10) and (11) respectively. Thus, the
population in the initial state $|f,0\rangle$ can be fast
transferred to that in the final state $|s,1\rangle$. Note that the
final state $|s,1\rangle$ contains only one photon in the field
which contains the coded information we want to transfer, therefore,
we should design an appropriate time interval between two atoms, and
make the second atom obtain the coded information from the cavity
field before the photon leaks out of the cavity.

\subsection{Ultrafast QST between two $\Lambda$-type atoms}

As emphasized in the previous section, we send two $\Lambda$-type
atoms through the cavity with a special time interval $\Delta T$, as
plotted in Fig. 2.
\begin{figure}
\includegraphics[width=0.8\columnwidth]{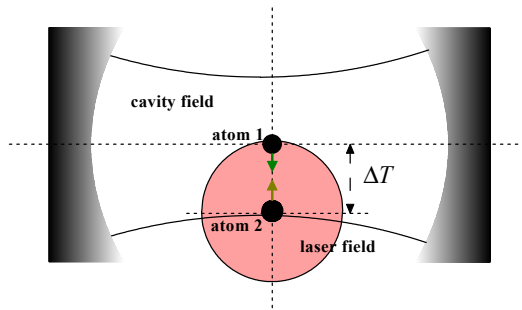}
\caption{Experimental setup for the ultrafast QST between two
$\Lambda$-type atoms through the cavity field. The atoms are denoted
by atom 1 and atom 2 respectively. Each solid arrow represents the
trajectory of each atom entering the cavity. The laser field is
confined within the red circle area. The special value $\Delta T$ is
chosen so that atom 2 can quickly obtain the coded information from
the cavity field when atom 1 interacts with the cavity field. The
dash line represents the reference coordinate axis.} \label {Fig.2}
\end{figure}

If the initial state for the atom-cavity system is $|fs,0\rangle$
(the symbol $|fs,0\rangle$ represents that atom 1 is in $|f\rangle$
state, atom 2 is in $|s\rangle$ state, and the cavity field is in
the vacuum state), the atom-cavity system will evolve in the
single-excitation Hilbert space: $\Gamma_{af}$ $\equiv$ $\{$
$|\phi_{1}\rangle$, $|\phi_{2}\rangle$, $|\phi_{3}\rangle$,
$|\phi_{4}\rangle$, $|\phi_{5}\rangle$ $\}$, with
\begin{eqnarray}\label{e16}
|\phi_{1}\rangle&=&|fs,0\rangle, |\phi_{2}\rangle=|es,0\rangle,\cr
|\phi_{3}\rangle&=&|ss,1\rangle,  |\phi_{4}\rangle=|se,0\rangle,
 |\phi_{5}\rangle=|sf,0\rangle.
\end{eqnarray}
The main task here is to fast transfer the population from the state
$|\phi_{1}\rangle$ to $|\phi_{5}\rangle$, leaving the cavity field
in the vacuum state. According to invariant dynamics of the
multi-mode driving between a single $\Lambda$-type atom and the
cavity mode in the previous section, we design a time-dependent
laser pulse to drive the atoms within the interaction time $0 \leq t
\leq t_{f}$ ($t_f=0.5us$) simultaneously. For the experimental setup
in Fig. 2, due to the relative position between the cavity and laser
fields (as plotted in Fig. 3), we can divide the whole
atom-cavity-atom interaction process into three subprocedures when
the atoms pass through the cavity and laser fields, i.e., $O_{1}$
($-0.5us\leq t\leq 0$), $O_{2}$ ($0\leq t\leq 0.5us$), and $O_{3}$
($0.5us\leq t\leq 1us$). Therefore, the interaction Hamiltonian
$H_{tot}$ of the atom-cavity system, under the rotating-wave
approximation, is: $H_{tot}$ $=$
$\sum_{l=1}^{2}[\Omega_{l}(t)|f\rangle_{l}\langle
e|+g_{l}(t)|e\rangle_{l}\langle s|a+H.c.]$.
\begin{figure}
\includegraphics[width=0.8\columnwidth]{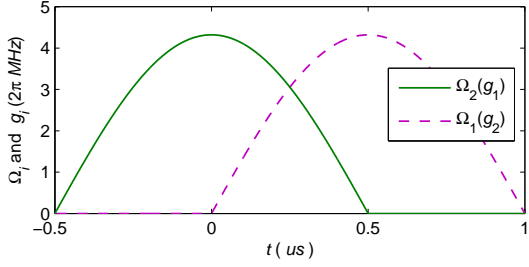}
\caption{The time-dependent coefficients $\Omega_i(t)$ and $g_i(t)$
($i=1,2$) versus $t$ for the corresponding geometry of the laser and
cavity fields. The parameters are chosen: $t_f=0.5us$ and
$\epsilon=0.1152$. The solid line and dash line represent the
parameters $\{\Omega_{2}(t),g_{1}(t)\}$ and
$\{\Omega_{1}(t),g_{2}(t)\}$, respectively.} \label {Fig.2}
\end{figure}

For the subprocedure $O_{1}$ ($-0.5us\leq t\leq 0$), atom 1 is sent
through the cavity and arrives at the center of the cavity field
when $t=0$, while atom 2 is sent through the cavity with a time
delay $\Delta T=0.5us$ and arrives at the center of the laser field
when $t=0$. Consider the initial state of the atom-cavity system is
$|fs,0\rangle$, the transition $|f\rangle\leftrightarrow|e\rangle$
in atom 1 is forbidden due to the absence of the laser field, and
the transition $|s\rangle\leftrightarrow|e\rangle$ in atom 2 is also
forbidden due to the absence of the photon in the cavity field.
Therefore, when $t=0$, the atom-cavity system keeps in the state
$|\psi(0)\rangle=|fs,0\rangle$.

For the subprocedure $O_{2}$ ($0\leq t\leq 0.5us$), atom 1
encounters the cavity and laser fields with the respective forms:
\begin{eqnarray}\label{e17-e18}
g_{1}(t)&=&(\pi/2t_f)\cot\epsilon\cos(\pi t/2t_f),\\
\Omega_{1}(t)&=&(\pi/2t_f)\cot\epsilon\sin(\pi t/2t_f),
\end{eqnarray}
while atom 2 encounters the cavity and laser fields with the
respective forms:
\begin{eqnarray}\label{e19-e20}
g_{2}(t)&=&(\pi/2t_f)\cot\epsilon\cos\big[\pi (t-\Delta T)/2t_f\big],\\
\Omega_{2}(t)&=&(\pi/2t_f)\cot\epsilon\sin[\pi (t+\Delta
T)/2t_f\big].
\end{eqnarray}
Then the atom-cavity system's evolution becomes:
\begin{eqnarray}\label{e21}
|\psi(t)\rangle&=&\sum_{k=1}^{5}D_{k}(t)|\phi_{k}\rangle,
\end{eqnarray}
where $D_{k}(t)$ is the time-dependent coefficient for the state
$|\phi_{k}\rangle$.

Under the laser pulse $\Omega_{1}(t)$ and the atom-cavity coupling
$g_{1}(t)$, the transition
$|\phi_{1}\rangle\rightarrow|\phi_{3}\rangle$ for atom 1 quickly
happens; at the same time, the transition
$|\phi_{3}\rangle\rightarrow|\phi_{5}\rangle$ for atom 2 also
quickly happens under $\Omega_{2}(t)$ and $g_{2}(t)$. The whole
transition $|\phi_{1}\rangle\rightarrow|\phi_{5}\rangle$ process
takes advantage of the cavity field as a medium for transporting the
coded information between two atoms. The population in the initial
state $|\psi(0)\rangle=|fs,0\rangle$ is transferred to that in the
final state $|\psi(t_{f})\rangle=|sf,0\rangle$. This process is very
different from that in Ref. \cite{PRA-86-033405-2012} where the
population transfer is limited within two internal states of a
single atom.

For the subprocedure $O_{3}$ ($0.5us\leq t\leq 1us$), the initial
state for the atom-cavity system is $|sf,0\rangle$, which does not
evolve with time $t$ due to the absences of photons in the cavity
field and the laser driving for atom 1 and atom 2, respectively.

If the initial state for the atom-cavity system is $|ss,0\rangle$,
the whole system under the cavity and laser fields keeps on the
state $|ss,0\rangle$ due to absence of photons in the cavity field.
Therefore, the fast QST between two atoms is achieved leaving the
cavity field in the vacuum state, i.e.,
$|s\rangle_{2}\bigotimes(x|f\rangle_1+y|s\rangle_1)$ $\rightarrow$
$(x|f\rangle_2+y|s\rangle_2)\bigotimes|s\rangle_{1}$ (the subscripts
$1$ and $2$ respectively denote atom 1 and atom 2), where $x,y$ are
the coded informations and $|x|^2+|y|^2=1$.

As we can see from the above analysis, the time evolution of the
initial state $|ss,0\rangle$ does not change during the whole
operation, thus its transfer keeps 100\%. The only factor that
affects the fidelity of our QST is the time evolution of the initial
state $|fs,0\rangle$. Therefore, to verify the reliability of our
QST between two atoms, we should consider the detailed
characteristic parameter of the atom-cavity system in the
subprocedure $O_{2}$ ($0\leq t\leq t_{f}$).
\begin{figure}
\includegraphics[width=0.7\columnwidth]{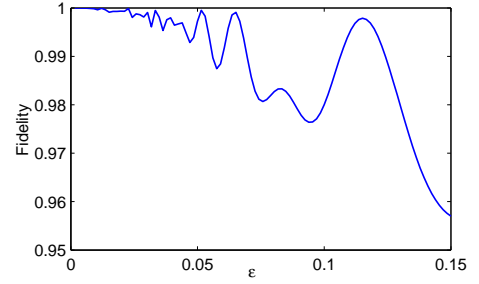}
\caption{ The fidelity $F$ vs the $\epsilon$.} \label {Fig.4}
\end{figure}

Although the analytic solution for the coefficient $D_{k}(t)$ is
hard to be obtained, we numerically verify the fidelity $F$ for the
final state $|\phi_{5}\rangle$ by setting the evolutive time
$t=0.5us$ in Fig. 4, where $F=|D_{5}(t)|^{2}$. Interestingly, for
the present multi-mode driving in the atom-cavity system, we find
that there are several distinct values where the fidelity turns out
to be close to unit, i.e., $\epsilon=0.1152$ for $N=1$;
$\epsilon=0.0651$ for $N=2$ and so on. In the following, we choose
the maximal value $\epsilon=0.1152$ to satisfy $F\simeq1$. The
behavior of $F$ against $\epsilon$ is oscillating, which is
essentially caused by different Lewis-Riesenfeld phases generated in
the transitions $|\phi_{1}\rangle\rightarrow|\phi_{3}\rangle$ and
$|\phi_{3}\rangle\rightarrow|\phi_{5}\rangle$ for the eigenvectors
of the invariant $I(t)$. This result coincides with that in the
single-atom system, meaning two atoms experience the similar
invariant dynamics of two internal states in a single $\Lambda$-type
atom \cite{PRA-86-033405-2012}. We remark that the whole interaction
time needed to complete the QST between two atoms is only $0.5us$,
which is an improvement compared with the previous QST without using
the dynamics of invariant-based engineering in the cavity QED system
\cite{PRA-71-023805-2005}.

\section{Analysis of experiment feasibility}
\begin{figure}
\centering\subfigure[]{\label{Fig.sub.b}\includegraphics[width=0.8\columnwidth]{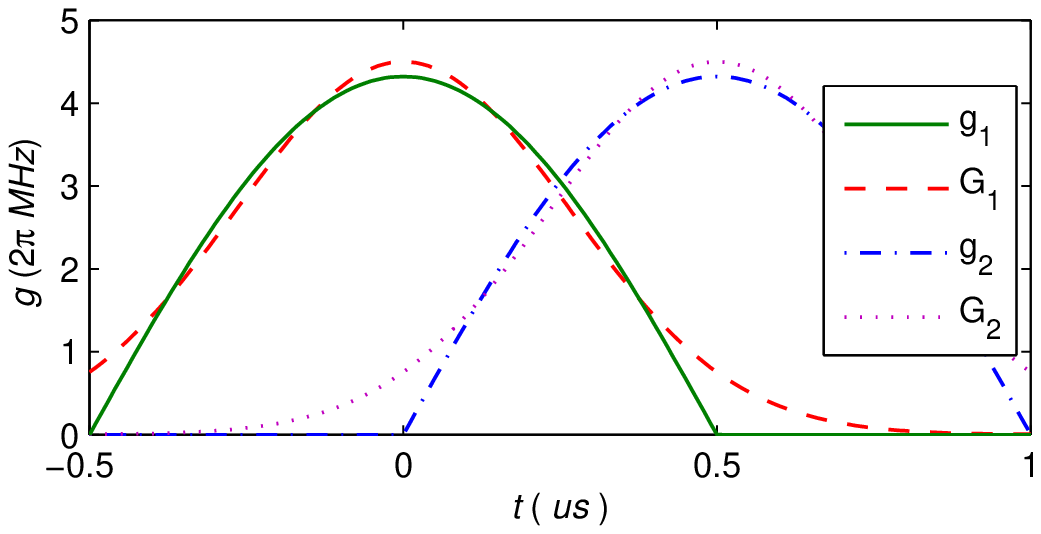}}
\subfigure[]{\label{Fig.sub.a}\includegraphics[width=0.8\columnwidth]{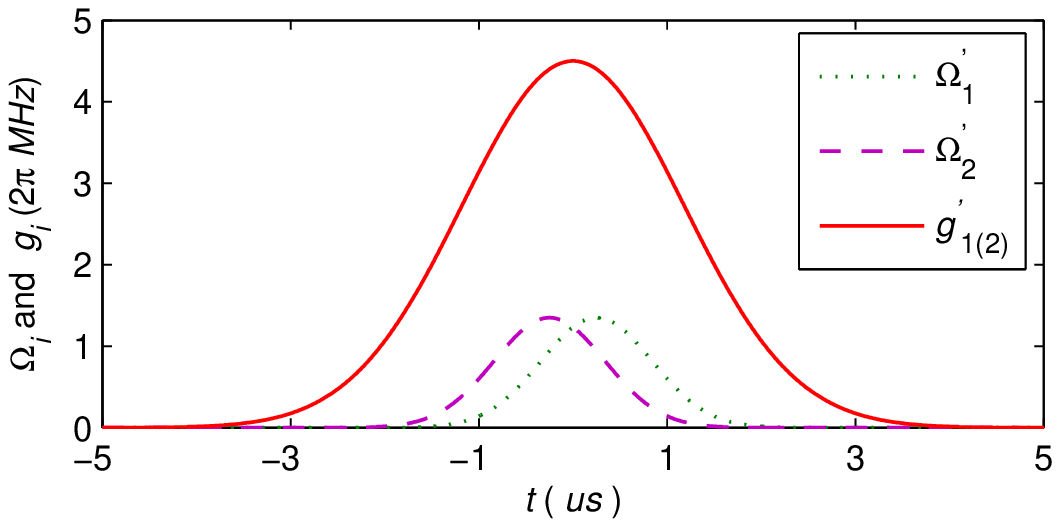}}
\caption{(a) The green solid and red dash lines represent the
atom-cavity coupling coefficients $g_{i}(t)$ ($i=1,2$) with the
sinusoidal-wave form and $G_{i}(t)$ with the redesigned
Gaussian-wave form, respectively. (b) The parameters of general
adiabatic passages $\Omega_i^{'}(t)$ and $g_{1(2)}^{'}(t)$ for the
traditional QST between two atoms in the cavity QED
system.}\label{Fig.3}
\end{figure}
To achieve the ultrafast QST between two $\Lambda$-type atoms, we
have obtained the analytic expressions of the time-dependent laser
pulse and the atom-cavity coupling strength, as depicted from Eqs.
(17) to (20), the shapes of which are sinusoidal-wave forms.
However, for the realistic experiment, it is more easy to obtain the
atom-cavity coupling strength $g(t)$ with the Gaussian-wave form.
Therefore, we use the mathematical method of minimum quadratic
fitting to redesign the wave forms for the parameters $g_{1}(t)$ and
$g_{2}(t)$; then we get the redesigned parameters $G_{1}(t)$ and
$G_{2}(t)$ with the Gaussian-wave forms as following:
\begin{eqnarray}\label{e22-e23}
G_{1}(t)&=&\epsilon^{'}\exp\big(\frac{t^2}{\sigma^2}\big),\\
G_{2}(t)&=&\epsilon^{'}\exp\big[\frac{(t-\Delta
T)^2}{\sigma^2}\big],
\end{eqnarray}
where $\epsilon^{'}=4.5\times2\pi MHz$ and $\sigma=\sqrt{0.14}us$,
as plotted in Fig. 5(a).

\begin{figure}
\centering\subfigure[]{\label{Fig.sub.a}\includegraphics[width=0.8\columnwidth]{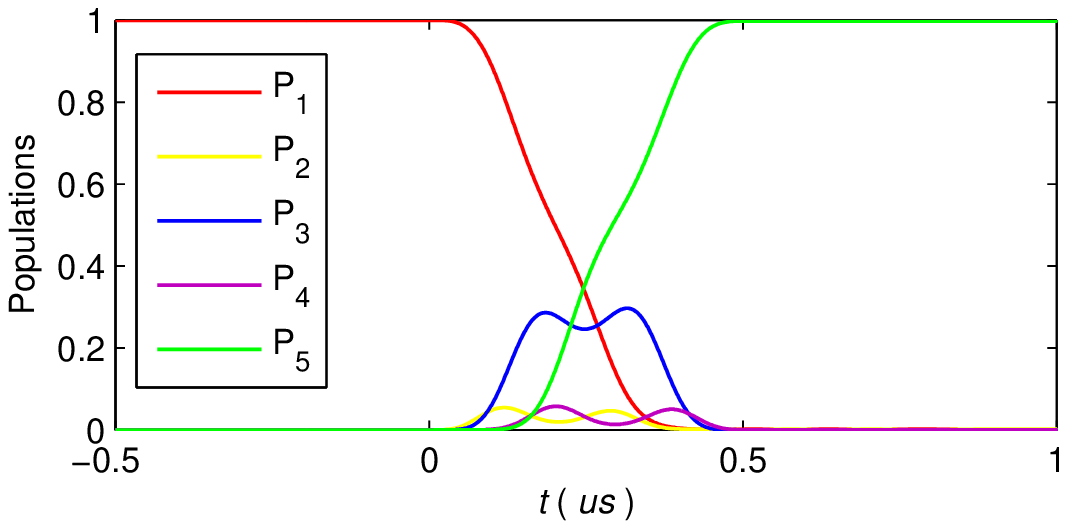}}
\subfigure[]{\label{Fig.sub.b}\includegraphics[width=0.8\columnwidth]{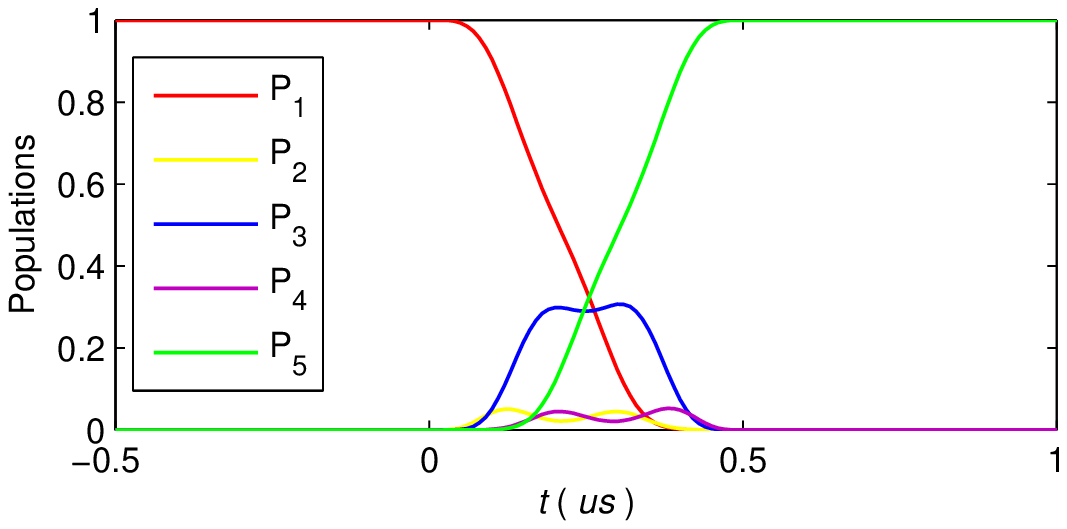}}
\subfigure[]{\label{Fig.sub.a}\includegraphics[width=0.8\columnwidth]{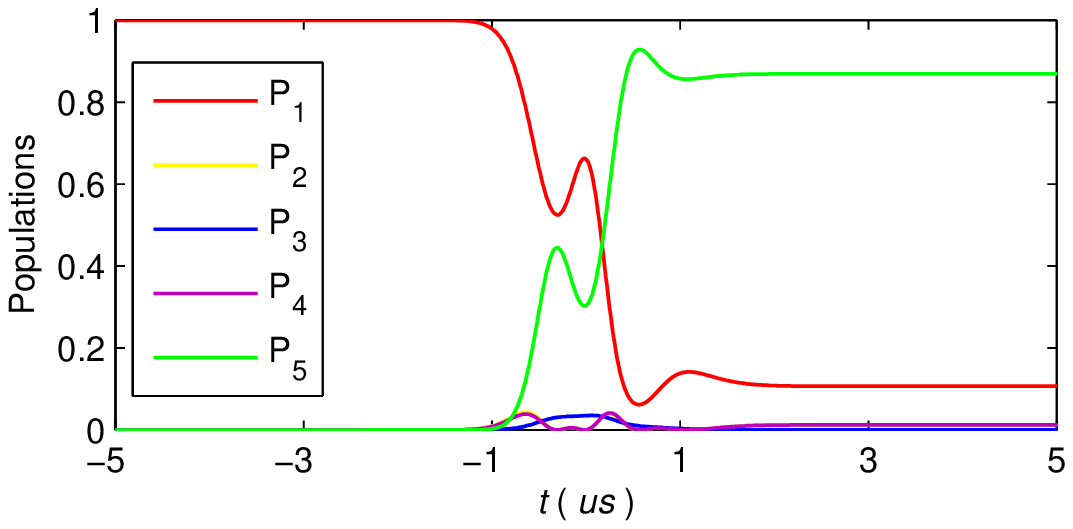}}
\caption{ Time evolution of the population $P_{k}$ ($k=1,2,3,4,5$)
with the set of parameters: (a) $\{$ $\Omega_{1}(t)$,
$\Omega_{2}(t)$, $g_{1}(t)$, $g_{2}(t)$ $\}$; (b) $\{$
$\Omega_{1}(t)$, $\Omega_{2}(t)$, $G_{1}(t)$, $G_{2}(t)$ $\}$; (c)
$\{$ $\Omega_{1}^{'}(t)$, $\Omega_{2}^{'}(t)$, $g_{1(2)}^{'}(t)$
$\}$.}\label{Fig.5}
\end{figure}

To verify the feasibility of the redesigned atom-cavity coupling
strength, we numerically simulate the population $P_{k}$ of the
state $|\phi_{k}\rangle$ by combining Eqs. (18), (20), (22), and
(23), in which $P_{k}$ is defined as $P_{k}=|D_{k}(t)|^2$, as
depicted in Fig. 6 (a) and (b). It is obvious to see that the
population $P_{k}$ by taking the redesigned atom-cavity coupling
strength $G_{i}(t)$ ($i=1,2$) with the Gaussian-wave form agrees
very well with that by taking the atom-cavity coupling strength
$g_{i}(t)$ with the sinusoidal-wave form, except for the slight
deviations between the corresponding excited states.

To compare with the traditional QST between two atoms based on the
adiabatic passage in the cavity QED system, we replace the present
set of parameters $G_{i}(t)$ and $\Omega_{i}(t)$ ($i=1,2$) with the
general set of parameters $g_{1(2)}^{'}(t)$ and $\Omega_{i}^{'}(t)$
\cite{PRA-72-012339-2005}:
\begin{eqnarray}\label{e24-e25-e26}
g_{1(2)}^{'}(t)&=&g^{'}\exp\big[\frac{(t-T_a/2)^2}{w_C^2}\big],\\
\Omega^{'}_{1}(t)&=&\Omega^{'}\exp\big[\frac{(t-T_a/2-d)^2}{w_L^2}\big],\\
\Omega^{'}_{2}(t)&=&\Omega^{'}\exp\big[\frac{(t-T_a/2+d)^2}{w_L^2}\big],
\end{eqnarray}
where $T_a=10us,\ w_C=\frac{T_a}{6},\ w_L=\frac{T_a}{12},\
g^{'}=4.5\times2\pi MHz,\ \Omega^{'}=0.3g^{'}$, and
$d=\frac{T_a}{40}$, as plotted in Fig. 5(b). Based on the parameters
$g_{1(2)}^{'}(t)$ and $\Omega_{i}^{'}(t)$, we numerically simulate
the corresponding population $P_{k}$ in Fig. 6(c). Apparently, the
total operation time in Fig. 6(c) needed to complete the QST is
rather longer than that in Fig. 6(a) or (b), and the population in
Fig. 6(c) drops below $90\%$, which is far smaller than that in Fig.
6(a) or (b). This is an obvious improvement both in the operation
time and the fidelity when we use the invariant-based inverse
engineering to implement the QST in the cavity QED system.
Furthermore, we also consider the effect caused by the fluctuations
of the redesigned parameters $\epsilon^{'}$ and $\sigma$ on the
fidelity $F$ in Fig. 7, and the result shows that the fidelity keeps
higher than $97\%$ even when the fluctuations $\delta \epsilon^{'}$
and $\delta \sigma$ are both $10\%$. Thus, the present QST using the
redesigned parameters $G_{1}(t)$ and $G_{2}(t)$ with the
Gaussian-wave forms is rather robust against the fluctuations of
$\epsilon^{'}$ and $\sigma$, and thus it is proved to be valid.

\begin{figure}
\includegraphics [width=0.8\columnwidth]{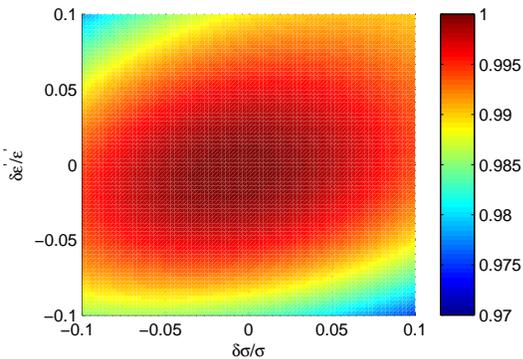}
\caption{The fidelity $F$ versus the fluctuations
$\delta\epsilon^{'}/\epsilon^{'}$ and $\delta\sigma/\sigma$.} \label
{Fig.7}
\end{figure}

Now we concentrate on discussing the fidelity $F$ in the presence of
dissipation caused by the noisy environment, which includes the
atomic spontaneous emission and the photon leakage out of the
cavity. When taking the effect of decoherence into account, the
master equation for the density matrix $\rho(t)$ of the present
atom-cavity system is expressed as:
\begin{eqnarray}\label{e27}
\dot{\rho}&=&-i[H_{tot},\rho]-\frac{\kappa}{2}(a^\dag a\rho-2a\rho
a^\dag+\rho a^\dag
a)\cr\cr&-&\sum^2_{k={1}}\sum_{m=s,f}\frac{\Gamma_{em}^k}{2}
(S_{em}^k S_{me}^k\rho-2S_{me}^k\rho S_{em}^k +\rho
S_{em}^kS_{me}^k),\cr&&
\end{eqnarray}
where $\Gamma_{em}^k$ is the atomic spontaneous emission rate from
the excited state $|e\rangle$ to the ground state $|m\rangle$
($m=s,f$) of the $k$th atom, $S_{em}^k$ $=$ $|e\rangle\langle m|$
and $S_{me}^k$ $=$ $|m\rangle\langle e|$ in the $k$th atom. $\kappa$
is the photon leakage rate. We assume $\Gamma_{em}^k=\Gamma/2$ for
simplicity. In Fig. 8 we plot the fidelity $F$ of the final state
$|sf,0\rangle$ versus the dimensionless parameters $\Gamma/g$ and
$\kappa/g$ via numerically solving the master equation (27) with the
set of parameter $\{$ $G_{1}(t)$, $G_{2}(t)$, $\Omega_{1}(t)$,
$\Omega_{2}(t)$ $\}$. The result of Fig. 8 shows that the QST
between two atoms based on the invariant-based engineering is
insensitive to the photon leakage and the atomic spontaneous
emission. This is because the present QST in the atom-cavity system
based on the invariant-based engineering is largely sped up, in
which the population from the initial state to the final state is
fast enough before the photon leaks out of the cavity, and the total
population for the excited states is rather small during the whole
system evolution. But, the fidelity of our QST is more sensitive to
the photon leakage than the atomic spontaneous emission because the
population in the excited states which involve one photon in the
field is far larger than that in the excited states which involve
the atomic excited state during the whole system evolution, i.e.,
$P_{3}\gg P_{2}+P_{4}$ in Fig. 6 (a) and (b).

\begin{figure}
\centering
\includegraphics[width=0.8\columnwidth]{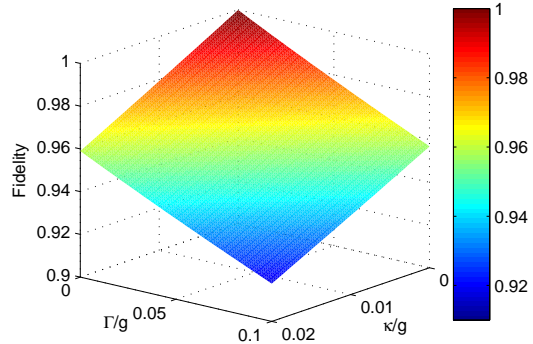}
\caption{The fidelity $F$ versus the dimensionless parameters
$\Gamma/g$ and $\kappa/g$.}\label{Fig.8}
\end{figure}
Finally, we present a brief discussion about the basic elements in
the real experiment. The $\Lambda$-type atomic configuration can be
achieved with the Cs atoms, in which the state $|f\rangle$
corresponds to $F=4, m=4$ hyperfine state of $6^2S_{1/2}$ electronic
ground state, $|s\rangle$ corresponds to $F=3, m=2$ hyperfine state
of $6^2S_{1/2}$ electronic ground state, and $|e\rangle$ corresponds
to $F=3, m=3$ hyperfine state of $6^2P_{1/2}$ electronic state. For
the typical experimental parameters $(g,\ \kappa,\
\Gamma)/2\pi=(750,\ 3.5,\ 2.62)\times MHz$, which have been reported
in the recent cavity QED experiments
\cite{PRA-71-013817-2005,PRA-67-033806-2003}, we find the fidelity
for the target state is still higher than $98.85\%$ with the set of
parameter $\{$ $G_{1}(t)$, $G_{2}(t)$, $\Omega_{1}(t)$,
$\Omega_{2}(t)$ $\}$. Therefore, the realization of the present QST
scheme is very promising with the current technology.

\section{CONCLUSION}

To conclude, we have proposed an alternative scheme to implement the
ultrafast quantum state transfer between two $\Lambda$-type atoms.
Compared with the traditional QST in cavity QED system, the present
QST based on the invariant-based inverse engineering is an obvious
improvement for both the operation time and the fidelity, in which
the quantum information is fast transferred between the atoms by
taking advantage of the cavity field as a medium. Through designing
the time-dependent laser pulse and atom-cavity coupling, we obtain
the target state with a high fidelity even when taking the atomic
spontaneous emission and the photon leakage out of the cavity. We
also redesign a Gaussian-type wave form in the atom-cavity coupling
for the realistic experiment operation.

\end{document}